\documentstyle[aps,prl,psfig,twocolumn]{revtex}
\begin{document}

\title{Disjoining Potential and Spreading of Thin Liquid Layers in the Diffuse Interface Model Coupled to Hydrodynamics}
\author{
Len M. Pismen$^{\it 1}$ and Yves Pomeau$^{\it 2}$\\
\small \it (1) Department of Chemical Engineering and Minerva
 Center for Nonlinear Physics of Complex Systems,\\ Technion--Israel Institute of Technology, Haifa 32000, Israel.\\
(2) Laboratoire de Physique Statistique de l'Ecole
Normale Sup\'erieure, associ\'e au CNRS,\\ 24 Rue Lhomond, 75231 Paris
Cedex 05, France.}
\date{\today}
\maketitle

\begin{abstract}
The hydrodynamic phase field model is applied to the problem of film spreading on a solid surface. The disjoining potential, responsible for modification of the fluid properties near a three-phase contact line, is computed from the solvability conditions of the density field equation with appropriate boundary conditions imposed on the solid support. The equation describing the motion of a spreading film are derived in the lubrication approximation. In the case of quasi-equilibrium spreading, is shown that the correct sharp-interface limit is obtained, and sample solutions are obtained by numerical integration. It is further shown that evaporation or condensation may strongly affect the dynamics near the contact line, and accounting for kinetic retardation of the interphase transport is necessary to build up a consistent theory.
\end{abstract}

\section{Introduction \label{intro}}

One of long-standing hydrodynamic riddles is the nature of viscous flow in the vicinity of a three-phase (gas-liquid-solid) contact line and the related problem of ``true'' and ``apparent'' dynamic contact angles  \cite{Duss79,DGen85}. The answer to the riddle must be, in fact, physico-chemical rather than purely hydrodynamic, since it depends on processes in the immediate vicinity of the three-phase boundary. The early detected paradox of a logarithmically divergent force required to displace the contact line \cite{HuSc} directly follows from the multivaluedness of the velocity field at the contact line -- if standard viscous hydrodynamics with a no-slip condition on the solid surface is to be believed. This paradox has been swept under a carpet rather than resolved by introducing a boundary condition with a stress or velocity dependent slip \cite{Duss76,hock76}. A drawback of hydrodynamic slip theories lies in their inherent inability to predict the dynamic contact angle. As a remedy, empirical relationships between the velocity and contact angle have to be introduced in model computations. 

Clearly, intermolecular forces, that determine the static contact angle to begin with, should have a say in a dynamic situation. Their direct action is, however, restricted to an immediate vicinity of the contact line, which is unobservable under available experimental resolution, so that an apparent contact angle seen at mesoscopic distances has to be strongly influenced by outer hydrodynamic conditions. Near the contact line itself, the properties of the fluid are different from those in the bulk, and even a common continuum description becomes questionable. 

Different approaches to description of the fluid motion in the vicinity of the three-phase boundary have been tried during the last two decades. The most straightforward way is to introduce intermolecular forces into the hydrodynamic equations of motion. This would lead, strictly speaking, to very difficult nonlocal equations, incorporating also the effects of variable density and diffuse interfaces \cite{keller}. Even in the sharp interface limit, a nonlocal dependence on the shape of the free interface leads to integro-differential equations which nobody as yet attempted to solve. A rational formulation is possible in lubrication approximation \cite{DGen85}, when the action of intermolecular forces reduces to a simple expression for disjoining pressure between parallel vapor-liquid and liquid-solid interphase boundaries \cite{Der}. This, however, does not eliminate the stress singularity, unless in the case of complete wetting when a sharp contact line is replaced by a gradual transition from a precursor film to a liquid film of macroscopic thickness \cite{DGen85}. At the same time, the usual expression for London--van der Waals forces leads to disjoining pressure divergent at small distances and necessitating a molecular-scale cut-off and leaving the ``true'' contact angle undetermined. This may be formally corrected by taking account of surface inclination \cite{hock93}, but the correction becomes effective at non-physical submolecular distances.

A radical solution is abandoning the continuum approach altogether in the immediate vicinity of the contact line. Slip is feasible on a microscopic scale where it may follow from activated diffusion of a first molecular layer \cite{druk}. Direct numerical simulations of molecular dynamics clearly demonstrate the effects of a diffuse boundary and effective slip at molecular distances \cite{Kop89,tr89}. Such simulations, however, cannot involve macroscopic volumes, and no ways to incorporate them in a macroscopic description are known. An alternative approach is to retain continuum description but to treat either 
vapor-liquid, or fluid-solid interface, or both as a separate phase with properties different from the bulk fluid. This approach was adopted by Shikhmurzaev \cite{shih} who relied also on deviation from thermodynamic equilibrium near the contact line as well as on the presence of a residual film to avoid the divergences and explain the difference between the static and dynamic contact angles. 

Treating the vapor-liquid interface as a separate phase may  be, indeed, justified  when surfactants are present, but otherwise a more natural way to account for its special properties of is to consider it as a region interpolating between the two phases. The origin of this approach is in the diffuse interface model going back to van der Waals himself \cite{vdw}. Much later, it became prominent in the phase field models \cite{cahn1}, used mostly in phenomenological theory of solidification where a fictitious phase field, rather than density, was used as a continuous variable changing across the interphase boundary. The theory of van der Waals was widely used for description of equilibrium fluid properties, including surface tension and line tension in three-phase fluid systems \cite{widom}. Applications of this theory to dynamical processes in fluids is much more difficult, as it requires coupling to hydrodynamics. The applicable equations were formulated rather recently \cite{jsvn,amcf}. Seppecher \cite{sepp} and Jacqmin \cite{jacq} solved the equations of the continuous density field coupled to the Stokes equation numerically in a small inner region near the contact line, matching it to the outer region where the standard sharp-boundary hydrodynamic limit applies. The prominent feature of the flow in the inner region was a substantial advective mass transport through the interphase boundary which served as an effective slip mechanism relieving the viscous stress singularity.

The aim of this communication is a rational analysis of the hydrodynamic phase field (diffuse interface) model based on the lubrication approximation. After formulating the basic equations in Section \ref{BASIC}, we reiterate the equilibrium relations defining the surface tension on all three kinds of interphase boundaries (Section \ref{SYL}) and discuss appropriate boundary conditions on the solid surface (Section \ref{S1bc}). This is followed by approximate computation of the density profile, equilibrium chemical potential and energy of the fluid layer (Sections \ref{S1dd}, \ref{S1ds}). The results of 1$d$ computations further serve as a basic ``vertical'' structure of the lubrication theory of Section \ref{2SCALE}, where a slow dependence on the ``horizontal'' coordinate is added. We show that the equations give a correct sharp-interface limit in both static and dynamic situation. The evolution equation derived in the lubrication approximation can be integrated numerically  yielding the dependence of the spreading velocity on a driving force. We shall see in Section \ref{sqeq} that no singularities develop in the case when the boundary condition fixes a unique fluid density at the solid surface.

This ``quasi-equilibrium'' theory is modified in Section \ref{NEQ} where a change of chemical potential across the fluid layer is taken into account. We start with discussing the ``vertical'' structure of chemical potential associated with viscous and kinetic retardation of steady motion of a vapor-liquid interface, and identify the dilute (vapor) phase as the locus of substantial variation of chemical potential. The potential drop is then computed numerically in Section \ref{krv}, yielding a relation between the disjoining potential and the flux across isodensity lines. This flux, which may be interpreted as incipient evaporation or condensation may help to alleviate viscous stress singularity when the boundary conditions make a sharp three-phase contact line necessary.

\section{Basic equations \label{BASIC}}

A general phase field model coupled to hydrodynamics includes the following elements: (1) a dynamic equation of the phase field variable(s) derived from an appropriate energy functional;  (2) a constituent relation defining the dependence of pressure or chemical potential on the phase variable(s); (3) the continuity equation; (4) the equation for a flow field ${\bf u}({\bf x},t)$. 

In a one-component system, the appropriate phase field variable is density $\rho$, General hydrodynamic equations for non-equilibrium systems with diffuse interphase boundaries are found in the recent review by Anderson {\it et al} \cite{amcf}.

The equation for the static density distribution is derived from the energy functional
\begin{equation}
{\cal F}= \int {\cal L}\, d^3{\bf x}, \;\;
{\cal L}=\rho f(\rho) +\mbox{$\frac{1}{2}$} K | \nabla\rho |^2 -\mu \rho,
\label{ff}  \end{equation}
where $\mu$ is the Lagrange multiplier (chemical potential) that serves to insure the mass conservation condition. The corresponding Euler -- Lagrange equation is
\begin{equation}
K \nabla^2 \rho -\partial_\rho (\rho f(\rho)) +\mu =0,
\label{phf}  \end{equation}
We shall suppose that the function $ f(\rho)$ is such that Eq.~(\ref{phf}) admits two stable solutions, $\rho=\rho_v$ and $\rho=\rho_l$, separated by an unstable solution $\rho=\rho_u$ and $\rho_v<\rho_u <\rho_l$. The solutions are at Maxwell construction (i.e.\ have equal energy $\rho f(\rho)$) at $\mu=0$, so that the chemical potential can serve as a bias parameter. 

The density field is coupled to hydrodynamics through the capillary tensor
 \begin{equation}
{\bf T} = {\cal L}{\bf I }-\nabla \rho \otimes \partial {\cal L}/\partial \nabla \rho  ,
\label{pp}  \end{equation}
where {\bf I}  is the unity tensor. Eliminating the Lagrange multiplier with 
the help of Eq.~(\ref{phf}) yields 
\begin{equation}
{\bf T} = (\mbox{$\frac{1}{2}$} K | \nabla\rho |^2 + K\rho \nabla^2 \rho -p ) {\bf I }- K \nabla \rho \otimes \nabla \rho  ,
\label{pp1}  \end{equation}
where the thermodynamic pressure is defined as $p=\rho^2 f'(\rho)$.

Neglecting the inertial effects, the flow is described by the generalized Stokes equation
\begin{equation}
\nabla \cdot ({\bf T + S} ) + {\bf F} =0,
\label{stok}  \end{equation}
where ${\bf F}=-\nabla V$ is an external force and ${\bf S}$ is the viscous stress tensor with the components
\begin{equation}
S_{jk} =\eta (\partial_j v_k +\partial_k v_j) + (\zeta - \mbox{$\frac{2}{3}$}\eta )\delta_{jk} \nabla \cdot {\bf v} ,
\label{visc}  \end{equation}
where $\eta, \zeta $ are dynamic viscosities (generally, dependent on $\rho$), and $v_j$ are components of the velocity field {\bf v}.
The system of equations is closed by the continuity equation
\begin{equation}
\rho_t+ \nabla \cdot (\rho {\bf v}) =0, \\
\label{cont}  \end{equation}
 
The Stokes equation (\ref{stok}) is rewritten using Eq.~(\ref{pp}) as 
\begin{equation}
- \nabla (p +V) + K  \rho \nabla \nabla^2 \rho + \nabla \cdot{\bf S} =0 .
\label{dpp}  \end{equation}
A more transparent equivalent form, which can be obtained directly from Eq.~(\ref{pp}), includes, instead of pressure, the chemical potential defined by Eq.~(\ref{phf}): 
\begin{equation}
- \nabla V -K\rho \nabla \mu  
+ \nabla \cdot(\eta \nabla {\bf v}) 
 +  \nabla[(\zeta +\mbox{$\frac{1}{3}$}\eta)\nabla \cdot {\bf v}] =0 .
\label{stokl}  \end{equation}

Further on, we shall compute the density and velocity field assuming that the characteristic macroscopic length $L^*$ of the flow field, as well as the scale of density variation in the tangential direction far exceed the characteristic thickness $(K/f^*)^{1/2}$ of the diffuse interface, where $f^*$ is a characteristic value of $ f(\rho)$. This ``thin interface'' approximation is apt to break down in the vicinity of the contact line, unless it is complemented by the ``lubrication'' approximation, which assumes a small angle between the (diffuse) interphase boundary and the solid surface. 
The applicability of this approximation depends as well on the boundary conditions at the solid surface.

\section{Equilibrium relations \label{ESYL}}
\subsection{Surface tension and Young--Laplace relation \label{SYL}}

Before approaching our main task of the analysis of motion in the vicinity of a three-phase boundary, it is necessary to clarify relevant properties of the dynamic phase field model for the basic case of a diffuse interface between semi-infinite phases. For a static interface, the phase field determines in a usual way the equilibrium surface tension \cite{vdw,widom}. The standard surface tension is defined as the energy per unit area of a flat interface separating two semi-infinite phases. Static solutions dependent only on the coordinate $z$ normal to the interface can be easily found by solving Eq.~(\ref{phf}). Rescaling the coordinate by the characteristic width of the diffuse interface, and denoting \begin{equation}
g(\rho)=\partial_\rho[\rho f(\rho)], 
\label{gf}  \end{equation}
we have
\begin{equation}
\rho''(z)- g(\rho) + \mu =0.
\label{phf0}  \end{equation}
The two static solutions are approached at $z \to\pm \infty$, and the boundary is static at $\mu=0$.

The interfacial energy is computed most easily by using as a dependent variable
the distortion energy  $T=\frac{1}{2} \rho_z^2$. Then Eq.~(\ref{phf0}) is rewritten as
\begin{equation}
T'(\rho)- g(\rho) +\mu =0,
\label{php}  \end{equation}
Integrating this proves that the distortion energy equals the potential energy at any point:
\begin{equation}
\mbox{$\frac{1}{2}$}\rho_z^2 = \rho f(\rho)-\mu\rho.
\label{vir}  \end{equation}
Using this ``virial theorem'', we compute
\begin{equation}
\sigma = \int_{-\infty}^\infty \rho_z^2 dz 
= \int_{\rho_v}^{\rho_l} \sqrt{2(\rho f(\rho)-\mu\rho)}d\rho.
\label{sig}  \end{equation}

Solid-fluid interactions are characterized by an appropriate boundary condition at the solid surface, as elaborated below. Generally, the density at the solid surface will be different in the vapor or liquid phase; we denote the respective values as $\rho_{sv}$ and $\rho_{sl}$. Accordingly, the ``liquid-solid'' or ``vapor-solid'' surface tension $\sigma_l$ or $\sigma_l$ is computed, respectively, by replacing one of the integration limits in Eq.~(\ref{sig}) by $\rho_{sl}$ or $\rho_{sv}$. 

In the vicinity of a critical point, the appropriate function, restricted to small deviations from the critical density $\rho_c$, is a cubic $g(\rho) = \rho-\rho_c - (\rho-\rho_c)^3$. Since our aim is a qualitative description of a system far from criticality involving the vapor phase with negligible density, we shall choose a shifted cubic 
\begin{equation}
g(\rho) =  \rho(1-2 \rho) (1-\rho), 
\label{grho}  \end{equation}
which is at Maxwell construction at $\mu=0$. Then $f(\rho) = \frac{1}{2}\rho (1-\rho)^2$ and the equilibrium surface tensions are computed as
\begin{eqnarray}
\sigma &=& \int_{0}^1\rho (1-\rho) d\rho =\mbox{$\frac{1}{6}$}, \cr
\sigma_l &=& \int_{\rho_{sl}}^1 \rho (1-\rho)  d\rho =
\mbox{$\frac{1}{6}$} (1-\rho_{sl})^2(1+2\rho_{sl}), \cr
\sigma_v &=& \int_0^{\rho_{sv}} \rho (1-\rho)  d\rho =
\mbox{$\frac{1}{6}$}\rho_{sv}^2 (3-2\rho_{sv}).
\label{sigq}  \end{eqnarray}
The first formula can be also obtained directly using the standard kink solution that approaches $\rho_v=0$ at $z \to \infty$ and $\rho_l=1$ at $z \to -\infty$:
\begin{equation}
\rho_0(z)= \left(1+e^z \right)^{-1}.
\label{skink}  \end{equation}
This solution may be, however, distorted in the vicinity of a solid wall. 

The expressions for $\sigma$ and $\sigma_\pm$ combine to the 
Young--Laplace formula 
\begin{equation}
\sigma_v - \sigma_l =\sigma \cos \theta,
\label{YL} \end{equation}
where $\theta$ is the ``standard'' contact angle that should be observed at distances much larger than the thickness of the transition layer, i.e. unity in the dimensionless units of Eq.~(\ref{phf0}). 

The Young--Laplace formula is a consequence of the Noether theorem applied to solutions of Eq.~(\ref{phf}).  Suppose that the solid surface is coincident with the $x$ axis ($z=0$) and $\rho(x,z)$ tends to $\rho_g$ at $z\to \infty$. Very far on the left ($x \to -\infty$) the vapor is close to the solid, so that $\rho (x,z)$ tends to a solution $\rho(z)$ of Eq.~(\ref{phf0}) with $\mu =0$, such that $\rho(0) = \rho_{sv}$ and $\rho \to \rho_v$ as $z \rightarrow \infty$. On the other end, for $x \to +\infty$, the liquid is close to the solid, that is $\rho(x,z)$ tends for $ x$ large positive and $z \ll x$ toward a solution $\rho(z)$ of Eq.~(\ref{phf0}) with $\mu =0$ and $\rho(0) = \rho_{sl}$ and $\rho \to \rho_l$ as $z$ becomes very large. For a given $x$, there is, however, a value of $z$, close to $\tan \theta$, such that there is a liquid-vapor interface and for $z \gg x \tan\theta$, $\rho$ becomes very close to $\rho_v$, as requested. If $\rho(x,z)$ satisfies these conditions, the liquid-vapor interface is inclined at the angle $\theta$ to the solid on scales much larger than the microscopic interface thickness (although this angle may change at a closer approach). 

The Young--Laplace formula follows from the invariance of the problem with respect to translations in the $x$ direction. Multiply Eq.~(\ref{phf}) (with $K$ rescaled to unity) by $\partial \rho/\partial x$, and integrate over $z$ from $z=0$ to $\infty$. This yields, after integrating by parts
\begin{equation}
\frac{d}{dx} \left\{ \int_0^\infty \left[\mbox{$\frac{1}{2}$}( \rho_x ^2 -\rho_z ^2) - \rho f(\rho)\right] dz \right\}=0.
\end{equation}
The braced expression is constant along the $x$ axis. This constant can be computed for $x$ very large negative and very large positive in the configuration just described. Equating the results, one gets:
$$\sigma _v = \sigma_l +  \int_{-\infty}^\infty \left[\mbox{$\frac{1}{2}$}
(\rho_0'(\zeta))^2 (\cos^2 \theta -\sin^2 \theta) -\rho f(\rho)\right] \frac{d\zeta}{\cos\theta}, $$
where $\zeta$ is the coordinate normal to the vapor--liquid interface, and $\rho_0'(\zeta)$ is the standard kink solution. The algebraic term reduces to $\frac{1}{2}(\rho_0'(\zeta))^2$ by Eq.~(\ref{vir}), and the final result is the Young--Laplace formula (\ref{YL}). The result is not influenced by possible deviations from the standard contact angle at a close approach to the solid surface, but, of course, hinges on the applicability of Eq.~(\ref{phf}). Since the actual inclination angle is apt to change at large distances due to external forces, such as gravity or dynamic pressure, the ``standard'' angle may be in fact unobservable at either small or large distances from the solid.

\subsection{Boundary conditions \label{S1bc}}

If the action of the solid on the density field is short-range (compared to the thickness of the diffuse interface), it can be accounted for by appropriate boundary conditions at the solid surface. The boundary conditions are usually assigned with the help of the Cahn construction \cite{cahn1,DGen85} balancing the distortion energy, distributed over a layer of same order of magnitude as the thickness of the diffuse interface, and the energy of 
fluid--solid interaction concentrated at the boundary. A more consistent way to arrive at the same boundary condition is to allow a non-vanishing variation of the density at the solid boundary $\delta \rho_s$ when the energy functional (\ref{ff}) is varied. In one dimension, this leaves, after integrating by parts, the boundary term $ \rho'(0) \delta\rho_s$. If the dependence of the fluid--solid interaction energy on the fluid density near the wall is expressed by a quadratic polynomial
\begin{equation}
\gamma(\rho_s) =\gamma_0 - \gamma_1 \rho_s+ \mbox{$\frac{1}{2}$}\gamma_2 \rho_s^2, 
\label{ensolid}  \end{equation}
the coefficient at $\delta \rho_s$ vanishes, provided
\begin{equation}
\gamma_1 -\gamma_2 \rho_s + \left. \rho'(z)\right|_{ \rho =\rho_s} =0.
\label{bcsolid}  \end{equation}
which is the boundary condition equivalent to that obtained through the Cahn construction (although the latter is expressed in an awkward integral form, including a radical with an  indefinite sign).

If one assumes that the solid--fluid interaction is short-range compared to the thickness of the diffuse vapor-liquid interface, it is likely prevail locally in the vicinity of a solid wall. This corresponds to the limiting case of very large $\gamma_1, \gamma_2$, when a simpler Dirichlet boundary condition $\rho=\rho_s$ is enforced on the solid surface. The range $\rho_v <\rho_s<\rho_l $ corresponds then to partial wetting.  

With the latter boundary condition and the cubic $g(\rho)$, the contact angle is
$\cos \theta= -1 +6\rho_s^2 - 4\rho _s^3 $,and is close to 0 or $\pi$ when $\rho_s $ is close, respectively, to 1 or $-0$. If $ \rho_s=1-a$ with $0  <a \ll 1$, we have $\theta = 2\sqrt{3 }a$. The contact angle is zero (complete wetting) at $ \rho_s \geq 1$. This ``standard'' angle has nothing to do with a ``true'' contact angle at the solid surface. The later is not defined at all in the diffuse interface theory, since different isodensity levels behave in a qualitatively different way as the solid surface is approached. The only level that hits the solid surface at the right angle is $\rho=\rho_s$; the levels with $\rho<\rho_s$ are asymtotically parallel, and those with $\rho>\rho_s$ antiparallel to the surface.

A more consistent way to derive the boundary condition is to start with a general expression for the energy of molecular interactions
\begin{equation}
{\cal F}= \int \int \rho({\bf x})\rho({\bf x}') V(|{\bf x}-{\bf x}'|)
d^3{\bf x} d^3{\bf x}'   
\label{eint}  \end{equation}
The mean-field energy functional (\ref{ff}) can be obtained from (\ref{eint}) assuming that the density changes on a characteristic scale far exceeding the range of the potential $ V(|{\bf x}-{\bf x}'|)$ and expanding $\rho({\bf x}') = \rho({\bf x}) + ({\bf x}-{\bf x}') \cdot \nabla\rho({\bf x}) + \ldots$. The algebraic term in the Lagrangian (\ref{ff}) is obtained in the zeroth order, and the distortion energy in the second order of the expansion. 
These expressions are modified when a solid boundary lies within the range of the interaction potential. The influence of the wall may be particularly strong in the standard case of Lennard--Jones interaction potential or a simplified expression $V \propto ({\bf x}-{\bf x}')^{-6}$ with a short-range hard-core cut-off, which gives the interaction energy diverging as $ z^{-3}$ with the distance from the solid surface. 
The diverging part of the energy may be taken as the surface energy potential that has to be minimized to obtain the density at the solid surface $\rho_s$. Under conditions when the bulk potential has two minima corresponding to low (vapor) and high (liquid) densities, the surface potential may also have two minima but the respective values, say, $\rho_{sl}$ and $\rho_{sv}$, would be, generally, different from the bulk values $\rho_{l}$ and $\rho_{v}$. This brings us to a Dirichlet boundary condition similar to that postulated above, but with the essential difference that two distinct values are allowed, and are likely to be chosen at the solid surface contacting, respectively, the liquid and vapor phase. Unlike the case when the surface density is unique, all isodensity levels in the range $\rho_{sv} \leq \rho \leq \rho_{sl}$ hit the solid surface.

The boundary condition (\ref{bcsolid}) also allows distinct density levels $\rho_{sv} \neq \rho_{sl}$ in the areas of the solid surface bordering either vapor of liquid. Assuming, for example, $0< \gamma_1 =a \ll 1, \; \gamma_2=0$, we have $\rho_{sv} \approx a, \; \rho_{sl}\approx 1+a $. It appears, however, quite unnatural that the main term in the density expansion fixes the density gradient rather than the density itself, so that
non-monotonic density profiles are forbidden in the above example and, on the contrary, enforced when $ \gamma_1$ is negative.

\subsection{Density profile in a thin layer \label{S1dd}}

The interaction between the solid surface and the interphase boundary can be computed most easily in the case when both surfaces are parallel and normal to the $z$ axis. The static solution $\rho(z)$ can be found by solving Eq.~(\ref{phf0}) subject to the appropriate boundary conditions at the solid wall. Solving the one-dimensional phase field equation in the form Eq.~(\ref{php}) is elementary; for the cubic $g(\rho)$, the exact solution is expressed in elliptic functions. Finding an approximate solution satisfying the boundary condition $\rho(0)= \rho_s=1-a$ with $|a| \ll 1$ is, however, more elucidating.

We construct the solution by perturbing a standard kink solution $\rho_0(z-h)$ centered at $z=h$, e.g. Eq.~(\ref{skink}) for the cubic $g(\rho)$. The actual solution is approximated to the zero order by the standard kink only when $\rho_0(-h)$ is sufficiently close to unity; thus, $h$ must satisfy the condition $h>\ln(1/a)$. The density profile is expanded in the small parameter $a$:
\begin{equation}
\rho= \rho_0(z-h)+a \rho_1(z;h)+ \ldots .
\label{expand}  \end{equation}
For the time being, we assume $\mu=0$. Then the first-order equation is
\begin{equation}
\rho_1''(z) + g'(\rho_0) \rho_1 =0,
\label{phf10}  \end{equation}
subject to the boundary condition 
\begin{equation}
\rho_1(0)=-1+a^{-1}[1-\rho_0(-h)] \approx -1+ \psi,
\label{bc1}  \end{equation}
where $\psi =a^{-1}e^{-h} \le 1$. 

Due to the exponential decay of interactions, the correction to the zero-order solution is actually of a higher order of magnitude everywhere except an $O(\ln a^{-1})$ vicinity of the wall, where $\rho_0$ is close to unity. On this interval, Eq.~(\ref{phf1}) can be replaced by the equation with constant coefficients
\begin{equation}
\rho_1''(z) -\rho_1(z) =0,
\label{phf1c}  \end{equation}
The solution decaying at $z \to \infty$ is
\begin{equation}
\rho_1 (z) = -e^{-z}\left(1- \psi \right). 
\label{rho1}  \end{equation}

\begin{figure}[tb]
\psfig{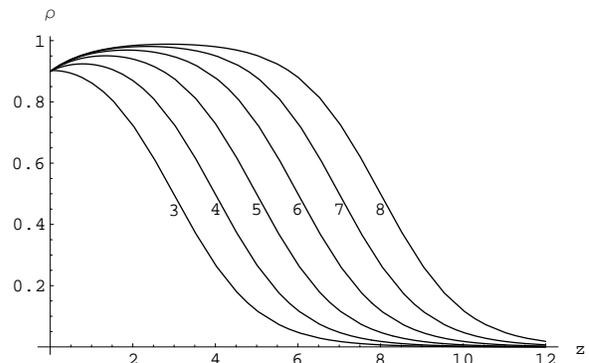}
\caption{Stationary density profiles. Numbers indicate the values of the nominal thickness $h$. \label{f0}}
\end{figure}

At $a>0, \: h>\ln(2/a)$, the combined function 
\begin{equation}
\rho_a =\rho_0+a \rho_1= \left(1+e^{z-h} \right)^{-1} - e^{-z}\left(a- e^{-h} \right)
\label{rhocom}  \end{equation}
reaches a maximum at $z=\frac{1}{2}\ln(ae^h -1)>0$ (Fig.~\ref{f0}). Such a solution describes a liquid layer sandwiched between the vapor and the solid. At smaller values of $h$, the maximum disappears, and the solution can be interpreted as a pure vapor phase thickening near the solid wall. The same solution applies at $a<0$ when the density increases at the solid surface, whether it is approached from the liquid phase or directly from the vapor phase.The approximation breaks down at $h<\ln(1/a)$, which is, in fact, below the minimal possible thickness of the dense layer in this model.

If the boundary condition allows two alternative fluid densities, a solution with $\rho(0)= a_v \ll 1$ may be also possible. This solution, corresponding to vapor phase adjacent to the solid surface, is simply $\rho \approx a_v e^{-z}$; this solution can be viewed as a tail of the basic kink centered at $z=\ln a_v<0$ (i.e.\ in the non-physical region). 

Non-monotonic density profiles are unstable. Since, however, the influence of the wall decays exponentially with the distance, the dynamics is practically frozen whenever the interphase boundary is separated from the wall by a layer thick compared to the characteristic width of the diffuse interface.

\subsection{Equilibrium chemical potential and energy \label{S1ds}}

A static solution with a fixed $h$ exists only at a certain fixed value of $\mu$, which can be determined using a solvability condition of the first-order equation. In a wider context, an appropriate solvability condition serves to obtain an evolution equation for the nominal position $h$ of the interphase boundary. The technique of derivation of  solvability conditions for a problem involving a semi-infinite region and exponentially decaying interactions is non-standard and therefore deserves special attention.

An inhomogeneous first-order equation has a general form
\begin{equation}
{\cal L} \rho_1 + {\cal H}(z) =0,
\label{phf1}  \end{equation}
containing an inhomogeneity ${\cal H}(z)$ and the linear operator 
\begin{equation}
{\cal L}= \frac{d^2}{ dz^2}+ g'(\rho_0). 
\label{phf1o}  \end{equation}
When Eq.~(\ref{phf1}) is defined on the infinite axis, the solvability condition of Eq.~(\ref{phf1}) appears due to the presence of an eigenfunction of ${\cal L}$ with zero eigenvalue related to the translational symmetry of the kink. The eigenfunction, obtained by applying the symmetry operator $d/dz$, is  simply $\rho_0'(z)$. The solvability condition is fixed by the orthogonality of the inhomogeneity to this eigenfunction:
\begin{equation}
\int_{-\infty}^\infty \rho_0'(z){\cal H}(z) dz =0.
\label{sc0}  \end{equation}
In the presence of a solid boundary, a difficulty arises, however, since the translational invariance is broken and no easily computable eigenfunction is available. In addition, the orders of magnitudes in the perturbative scheme should be estimated in a non-standard way in view of the exponential decay of interactions.

The difficulties are overcome with the help of asymptotic matching technique similar to that employed in the theory of vortex dynamics \cite{vort}. The solvability condition is computed, similar to Eq.~(\ref{sc0}), using the translational eigenfunction on the infinite axis, but the integration is not carried out over the entire axis (which now extends into the unphysical region $z<0$), but starts at some location $z=z_0>0$ where $\rho$ differs from the asymptotic value $\rho=1$ by an $O(a)$ increment. This generates boundary terms in the solvability condition, which takes now the form
\begin{eqnarray}
&& \int_{z_0}^\infty \frac{d\rho_0(z-h)}{dz}\,{\cal H}(z) dz = \cr
&& \left[ \frac{d\rho_0(z-h)}{dz}\, \frac{d\rho_1(z)}{dz} -
\frac{d^2\rho_0(z-h)}{dz^2}\, \rho_1(z) \right]_{z=z_0}.
\label{sc1}  \end{eqnarray}

The boundary values of the first-order solution $\rho_1'(z)$ are obtained by solving the first-order equation (\ref{phf1}) directly on the interval $0 \le z \le z_0$, where Eq.~(\ref{phf1}) can be replaced by the equation with constant coefficients (\ref{phf1c}) with the added inhomogeneity ${\cal H}(z)$. The solution of this equation is
\begin{equation}
\rho_1 (z) = \rho_1^{(h)} (z) + \int_0^z {\cal G}(z-\zeta){\cal H}(\zeta) d\zeta,
\label{rho1in}  \end{equation}
where $\rho_1^{(h)}$ is given by Eq.~(\ref{rho1}) and ${\cal G}(z-\zeta)$ is Green's function of Eq.~(\ref{phf1c}). The last term can be neglected for certain inhomogeneities, provided the lower limit of the integral in the left-hand side of Eq.~(\ref{sc1}) can be shifted to $-\infty$ without introducing a significant error. The matching is successful when Eq.~(\ref{sc1}) reduces to a form independent of $z_0$ in the leading order.

The simplest application of the above matching technique is the computation of a constant value of chemical potential $\mu=\mu_c$ required to keep the kink at equilibrium (possibly, unstable) at a given location $z=h$. In this case, the inhomogeneity in Eq.~(\ref{phf1c}) is just a constant ${\cal H}=\mu_c$, and the integral in the left-hand side of Eq.~(\ref{sc1}) is $\mu_c[\rho_0(\infty)-\rho_0(z_0)]= -\mu_c +O(a)$. Since this expression remains unchanged in the leading order when $z_0$ is shifted to $-\infty$, i.e. $\rho_0(z_0)=1-O(a)$ replaced by unity, it is sufficient to use in Eq.~(\ref{sc1}) the first term of Eq.~(\ref{rho1in}) only. Retaining the leading term only, we obtain 
\begin{equation}
\mu_c \equiv a^2 M(h) =  2a^2 \psi (1- \psi) = 2 e^{-h}\left( a- e^{-h}  \right).
\label{dplam}  \end{equation}
The first expression demonstrates that the computed chemical potentials in fact at most of $O(a^2)$, although the equation is nominally of the first order. The gained order of magnitude is due to the fast decay of interactions. Since the computed value is of a higher order, there is no need to correct the equilibrium profile computed in the preceding subsection to $O(a)$. For $a>0$, the function $\mu_c(h)$ passes a maximum at the same value $h=\ln(2/a)= O(1)$ that marks the transition from monotonic to non-monotonic density profiles. Sustaining a static profile requires a bias in favor of the liquid state, and the value of $\mu_c$ at the maximum represents the critical value of chemical potential required to nucleate a thick liquid layer on the solid surface. For $a<0$, $\mu_c$ in Eq.~(\ref{dplam}) is negative and increases monotonically  with $h$; in this case, on the contrary, a bias in favor of the vapor phase is necessary to keep the interface stationary.

For the boundary condition $\rho'(0)=- a$, the chemical potential of the dense solution is equivalent to Eq.~(\ref{dplam}) with the inverted sign of $a$. The rescaled value is
\begin{equation}
M(h)  = - 2 a^{-1}e^{-h}\left( 1+ a^{-1}e^{-h}  \right).
\label{dplam1}  \end{equation}

The correction to energy, defined as
\begin{equation}
E(h) =\int_{0}^{\infty}
 \left[ \rho f(\rho) +\mbox{$\frac{1}{2}$} \rho_z^2 \right] dz
  = \sigma +a^2 \widetilde E(h), 
\label{e1}  \end{equation}
also turns out to be of $O(a^2)$. The ``virial theorem'' used in Eq.~(\ref{sig}) does not hold to this order when density is defined by the first-order function $\rho_a$. The best way to compute the energy is to use directly the variational formulation to relate it with the computed chemical potential. Requiring the one-dimensional energy functional (\ref{ff}) to be extremal with respect to $h$ and using in the last term $\rho(z)= \rho_0(z-h)$, we compute
\begin{eqnarray}
\frac{d\widetilde E}{dh}=- M \int_{0}^{\infty}\rho_0'(z-h) \, dz = M +O(a).
\label{dpcx3} \end{eqnarray}

\section{Motion in a thin layer  \label{2SCALE}}

\subsection{Double-scale expansion \label{curved}}

Two-dimensional motion can be rationally treated in the familiar ``lubrication approximation'', assuming the characteristic scale in the ``vertical'' direction (normal to the solid surface) to be much smaller than that in the ``horizontal'' (parallel) direction. When the interface is weakly inclined and curved, the density is weakly dependent on the coordinate $x$ directed along the solid surface. Respectively, the vertical velocity $v$ is assumed to be much smaller than the horizontal velocity $u$. The scale ratio is determined by the contact angle, and should be set at $O(a)= O(\sqrt{\delta})$ to match the scaling of the phase field. The velocities $v,u$ corresponding to weak disequilibrium of the phase field considered above will be consistently scaled if one assumes $\partial_z=O(1), \; \partial_x=O(\sqrt{\delta}), \;  u=O(\delta^{3/2}), \; v=O(\delta^{2})$. It is further necessary for consistent scaling of the hydrodynamic equations that the ``constant'' part of the chemical potential $\mu$, associated with interfacial curvature, disjoining potential, and external forces and weakly dependent on $x$, be of $O(\delta)$, while the ``dynamic'' part varying in the vertical direction and responsible for motion across isodensity levels, be of $O(\delta^2)$. Further in this Section, we shall assume therefore that $\mu +V$ is independent of $z$; this assumption will be
re-examined in Section \ref{NEQ}.

In two dimensions, the term $\rho_{xx}$ is added to the inhomogeneity in the first-order equation (\ref{phf1}). In this order, the vertical density profile can be represented by the standard kink solution $\rho_0(z-h(x,t)) $, and the $x$ dependence is due to slow variation of $h$ in the ``horizontal'' direction. Thus, 
\begin{equation}
\rho_{xx}= - \rho_0'(z-h) h_{xx}+ \rho_0''(z-h) h_{x}^2.
\label{rxx}  \end{equation}
The respective contribution to the solvability condition is, in the leading order,
\begin{equation}
-h_{xx}\int_{-\infty}^\infty [\rho_0'(z)]^2 dz= -\sigma h_{xx},
\label{rxxi}  \end{equation}
while the contribution of the term containing $h_{x}^2$ vanishes in the leading order by symmetry. 

Another possible contribution to the solvability condition may come from external forces. In the presence of gravity directed against the $z$ axis, the equilibrium is achieved, according to Eq.~(\ref{stokl}), at $\mu = \mu_0 - a^2 Gz$ rather than $\mu = \mu_0=$ const. The rescaled acceleration of gravity is denoted as $ a^2 G$, which presumes that it matches the other terms by the order of magnitude. The integral in Eq.~(\ref{sc1}) involving the variable part of $\mu$ is mostly accumulated in the diffuse interface region, so that we have in the leading order
\begin{equation}
-G \int_{0}^\infty z \rho_0'(z-h) dz \approx  Gh·.
\label{rgi}  \end{equation}
Collecting Eqs.~(\ref{rxxi}) and (\ref{rgi}), we obtain the expression for the hydrostatic chemical potential
\begin{equation}
\mu = \delta \left[ M(h) - \sigma h_{xx} + G·( h - z ) \right],
\label{dpcx}  \end{equation}
where $ M(h)$ is defined by Eq.~(\ref{dplam}) or (\ref{dplam1}).

\subsection{Statics in lubrication approximation  \label{Statlubr}}

Equation (\ref{dpcx}) will be used later on to investigate dynamical processes where motion of the contact line is involved. In this subsection we investigate  the statics of this lubrication approximation, and show how it relates to the general Young--Laplace result on the static contact angle. It seems to be important  for the general consistency of the theory to have dynamical equations for the contact angle that reduce to the usual equilibrium theory in the absence of motion. In most realistic cases, the effect of gravity is negligible near the contact line, since gravitational forces are much weaker than molecular forces. Therefore, the statics of the contact angle, at scales in between molecular length scales and the capillary length (that is the length scale beyond which gravity plays a role), depends on solutions of Equation (\ref{dpcx})  without the gravity term $G·( h - z ) $. Moreover, as we want to study equilibrium situations where a liquid-vapor interface merges with the solid surface, the chemical potential $\mu$ is set to its equilibrium value, $0$, so that the equation under consideration is:
\begin{equation}
M(h) - \sigma h_{xx} =0.
\label{dpcx2}  \end{equation}

In order to derive from this equation the Young--Laplace condition, one can use the relation (\ref{dpcx3}) between the energy and chemical potential computed in the end of Section \ref{S1ds}. We integrate Eq.~(\ref{dpcx2}) with the boundary conditions for the function $h(x)$ such that for $x \to -\infty$, the vapor is close to the solid, while for $x \to \infty$, the liquid is close to the solid, until a height $h(x) \approx \theta x$, $\theta  \ll 1 $ where a liquid-vapor interface is situated. The relevant first integral of Eq.~(\ref{dpcx2}) reads:
\begin{equation}
\mbox{$\frac{1}{2}$}\sigma h_{x}^2 = \widetilde E (h) - \widetilde E(h_0),
\label{dpcx3a}  \end{equation}
where $ h_0$ is the root of $M(h)$ that gives the thickness of the precursor film lying between the solid and the vapor phase; $h_0 = \ln(1/a)$ in the model with a cubic $f(\rho)$ and Dirichlet boundary condition. The structure of Eq.~(\ref{dpcx3a}) is obviously similar to the Young--Laplace formula. The capillary energy at very large negative $x$, $\widetilde E (h_0)$, is nothing but the solid-vapor surface tension, $\sigma_v$. The capillary energy at very large positive $x$ where the vapor-liquid interface is far removed from the solid is the sum of the independent contributions of the solid-liquid and a free liquid-vapor interfaces. Integrating up to very large positive $x$, where $h\approx x \theta$, one gets therefore $\widetilde E (\infty) = \sigma + \sigma_l$. Using this in Eq.~(\ref{dpcx3a})  and subtracting $\sigma$ from both sides yields
\begin{equation}
\sigma -\mbox{$\frac{1}{2}$} \sigma \theta^2 \approx \sigma \cos\theta =  \sigma_v -  \sigma_l,
\label{dpcx4}  \end{equation}
which is the sought after Young--Laplace condition, derived from the equations of the lubrication approximation for the position of the liquid-vapor interface.

\subsection{Equations of motion in lubrication approximation  \label{lubr}}

The horizontal velocity $ u$ is determined from the horizontal component of the Stokes equation. Adding gravity as an external force, we write the leading order equation as
\begin{equation}
- \rho_0(z-h) P_x + ( \eta u_{z})_z  =0,
\label{stox}  \end{equation}
where the effective pressure $P$ is defined as
\begin{equation}
P= G \alpha x + M(h) - \sigma h_{xx} + G·( h - z ),
\label{stoxp}  \end{equation}
This expression follows from Eq.~(\ref{dpcx}), with the addition of the gravity term acting when the supporting plane is weakly inclined. The inclination angle $\alpha$ must be of $O(\sqrt{\delta})$ to match by the order of magnitude the other terms in the equation. The density profile is given in the leading order by the standard kink solution (\ref{skink}) centered at the nominal interface position $h(x)$ slowly varying in the horizontal direction. 

The solution of Eq.~(\ref{stox}) satisfying the no-slip boundary condition on the solid boundary and the no stress condition at infinity has a general form
\begin{equation}
u (z) =  \eta^{-1}P_x \Psi(z;h). 
\label{us} \end{equation}
The function $\Psi(z;h)$ depends on an assigned dependence of viscosity on density, but the flux $u\rho_0$ in the dense layer (at $z$ not much larger than $h$) is nearly the same for either $\eta = $ const or $\eta \propto \rho $, and is close to the standard lubrication solution $\Psi = - z(h -\frac{1}{2}z)$ valid for incompressible Poiseuille flow in a layer of thickness $h$ with a free boundary.

\begin{figure}[tb]
\psfig{figure=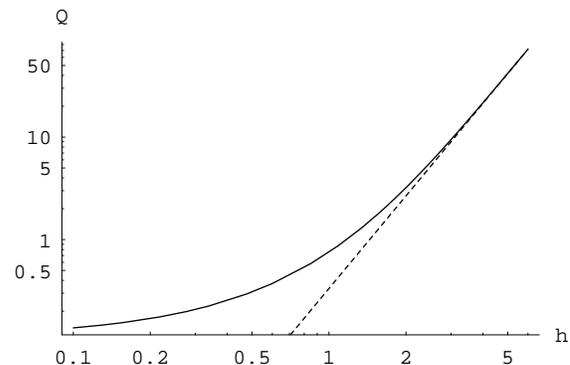,width=8.1cm}
\caption{The function $Q(h)$, compared with the respective function for the sharp interface $Q_0(h)=\frac{1}{3}h^3$ (dashed line) \label{f3}}
\end{figure}

The evolution equation of $h$ is obtained by inserting Eqs.~(\ref{skink}), (\ref{us}) in the continuity equation (\ref{cont}) and integrating it from 0 to $\infty$. Using the relations
\begin{eqnarray*}
 \int_0^\infty \rho_t dz =- h_t \int_0^\infty \rho_0'(z) dz &=& h_t + O(a), \cr
\int_0^\infty (\rho v)_z dz &=&0,
\end{eqnarray*}
we obtain
\begin{equation}
h_t = \eta^{-1}\partial_x \left[ Q(h) P_x \right].
\label{hs}  \end{equation}
where
\begin{equation}
Q(h) = - \int_0^\infty \rho_0(z-h) \Psi (z,h) \, dz.
\label{hsq}  \end{equation}
The function $Q(h)$, computed numerically and plotted in Fig.~\ref{f3}, differs only slightly from the respective function for the sharp interface $Q_0(h)=\frac{1}{3}h^3$ when $h$ exceeds its minimal admissible value $h_0=\ln(1/a)$. Taking into account small deviations from the standard kink solution near the wall adds only a higher-order correction.

\subsection{Quasi-equilibrium spreading \label{sqeq}}

Apart from a slightly modified volumetric rate, the specific contribution of the diffuse interface to Eq.~(\ref{hs}) is carried by the function $M(h)$, which is dependent on the boundary conditions on the solid surface and expresses disjoining potential. It should be emphasized that this function is not given {\it a priori} but computed in the framework of the phase field theory (Section \ref{S1ds}). The structure of Eq.~(\ref{hs}) is identical to that of standard equations of motion of thin liquid films, which are recovered at large $h$ when the disjoining potential becomes negligible. At small $h$, the disjoining potential is not singular as in the sharp-interface theories with van der Waals interactions \cite{DGen85}. At the same time, the viscous stress singularity at the contact line is relaxed as the latter's location becomes indefinite.   

Steady flow of a liquid film under the action of disjoining potential and gravity can be described by Eq.~(\ref{hs}) rewritten in the frame moving with a speed $U$. We shall assume that the liquid layer thickens at $x \to \infty$, and assume $U$ to be positive when the thick layer advances. Standard macroscopic arrangements fixing the asymptotic conditions at $x \to \infty$ are possible, e.g. $ h \to \infty, \; h_x = -\alpha$ for a liquid wedge with the angle $\alpha$ or $ h_x = 0, \; h=\sqrt{3U/\alpha G}$ for an asymptotically flat film on an inclined plane. 

Admissible asymptotics at $x \to -\infty$ depends on the form of the function $M(h)$. If it is given by Eq.~(\ref{dplam}) with $a>0$, the layer may attain asymptotically at $x \to -\infty$ the state of lowest energy $h=h_0=\ln(1/a)$ (formally, this is possible at zero inclination $\alpha$, although gravity effects are negligible in films of molecular thickness). 

The starting point is Eq.~(\ref{hs}) with the effective pressure given by Eq.~(\ref{stoxp}).  Removing extra parameters by rescaling and integrating once yields
\begin{equation}
h'''(x) - \left(M'(h)+ G \right)h'(x) - \alpha G + \frac{U(h-h_0)}{Q(h)}=0,
\label{hs1}  \end{equation}
where the integration constant has been introduced allowing for a precursor film with the thickness $h_0$ at $x \to -\infty$. A more convenient form of Eq.~(\ref{hs1}) is obtained using as the dependent variable $y=h_x^2=2T$ and as the independent variable the nominal thickness $h$:
\begin{equation}
\frac{1}{2}y''(h) - \left(M'(h)+ G \right) 
+ \frac{1}{\sqrt{y}} \left( \frac{U(h-h_0)}{Q(h)}- \alpha G \right) =0.
\label{hsy}  \end{equation}

\begin{figure}[tb]
\psfig{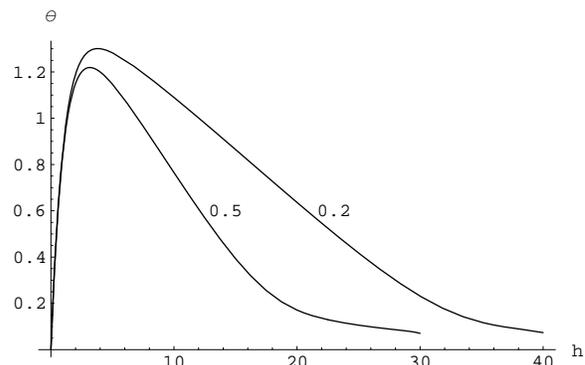}
\caption{Dependence of the interface inclination angle $\theta$ on the nominal thickness $h$ of a spreading dense layer for $3U=0.5$ and $3U=0.2$ (as indicated by numbers at the respective curves). The values of $G$ found by shooting are, respectively, 0.035123081 and 0.0079817. \label{fx}}
\end{figure}

Equation (\ref{hsy}) is free from singularities which are usually caused by divergences of either viscous stress, or disjoining potential, or both, in a layer of vanishing thickness. It can be integrated numerically starting from the asymptotics at $x \to -\infty$. The asymptotics of Eq.~(\ref{hsy}) obtained by expanding near $h=h_0$ is $y \asymp c^2 (h-h_0)^2$, implying exponential decay to the ``optimal'' thickness $ h - h_0 \propto e^{\kappa x}$, where the constant $\kappa$ is a positive root of the characteristic equation 
\begin{equation}
\kappa^3 - M'(h_0) \kappa+ U/Q(h_0) =0.
\label{hsyc}  \end{equation}
Fixing, say, the value of $U$, one can use the shooting method to adjust the value of $G$ satisfying the appropriate boundary condition at infinity, $\sqrt{y}=-\alpha$. A very fine adjustment of the parameter is needed to advance to moderate values of $h$. An example of a computed dependence of the interface inclination angle on the nominal thickness of a dense layer spreading on a horizontal support is shown in Fig.~\ref{fx}.

\section{Non-equilibrium motion  \label{NEQ}}

\subsection{Viscously retarded motion  \label{ret}}

Equilibrium solutions with $\rho$ varying along the $z$ axis exist only at a particular constant value of $\mu$, equal to zero in the adopted gauge. Any deviation of this value sets the interface into motion; the interface shift corresponds to evaporation or condensation retarded by viscous friction. The simplest case is  steady propagation of the boundary between two semi-infinite phases. The stationary one-dimensional equations in the frame moving with the speed $c$ of the steadily propagating interface are
\begin{equation}
 (\rho  v)_z =0, \;\;\;  -\rho \mu_z +( \widehat\eta v_{z})_z  =0
\label{cont1d}  \end{equation}
where $v$ is the single velocity component in this frame; external forces are omitted and $\widehat \eta=\zeta +\frac{4}{3}\eta $ is the renormalized viscosity, accounting also for the divergence term in Eq.~(\ref{stokl}).  These equations are readily integrated yielding
\begin{equation}
 j \equiv \rho  v = \mbox{const},  \;\;\; 
 \mu=\mu_c + j R(z), 
\label{lam1d}  \end{equation}
where
\begin{equation}
R(z)= \int \frac{1}{\rho}\frac{d}{dz}
  \left(\widehat \eta \frac{d\rho^{-1}} {dz}\right) dz.
\label{lam1dr}  \end{equation}
The flux $j$ is related to the propagation velocity $c$ as $j= -c(\rho_l -\rho_v)$. The sign of $c$ is chosen in such a way that it is positive when the dense (liquid) state advances. The constant $\mu_c$, which may be fixed by external conditions, represents the driving force of the process.

It is reasonable to assume that the disequilibrium is weak, so that both $\mu_c$ and the constant flux $j$ are multiplied by a book-keeping small parameter $\delta$ when Eq.~(\ref{lam1d}) is used in Eq.~(\ref{phf0}). The perturbed equation can be expanded in a usual way, and the relation between the flux $j$ and $\mu_c$ is obtained from the solvability condition (\ref{sc0}): 
\begin{equation}
\mu_c  = - c \int_{-\infty}^\infty  \rho_0'(z) R(z) dz .
\label{lam1s}  \end{equation}

The integral in the right-hand side can be interpreted as the effective friction factor. It depends on the basic density profile $\rho_0(z)$ as well as on the assumed dependence of the viscosity on density. If $\rho_0=\rho_c+\widetilde \rho$ represents a weakly perturbed critical density, $R(z)= -\eta_c  \rho_c^{-3}\widetilde \rho_z$, and the integral in Eq.~(\ref{lam1s}) is proportional to surface tension. In the case of vanishing vapor density which interests us most, assuming $\eta=$ const leads to a divergent integral.
The divergence is not eliminated  also when the viscosity is proportional to density. Taking, for example, $\widehat \eta=\nu \rho$, Eq.~(\ref{lam1dr}) is evaluated using the relation $\rho_z = -\rho(1-\rho)$ as $R(z)= - \nu \ln (\rho_0(z)/ \rho_c)$. The weak divergence on the vapor side can be eliminated by assuming a small but finite vapor density $\rho_v$. Then evaluating the solvability condition yields 
\begin{eqnarray}
\mu_c &=&- c \int_{-\infty}^\infty \rho_0'(z) R(z) dz \cr &=& 
-c \nu \int_{\rho_v}^{\rho_l} \ln \frac{\rho_0(z)}{ \rho_c} d \rho =
 - c \nu (1+\ln \rho_c).
\label{muinf}  \end{eqnarray}
where $\mu_c$ is the chemical potential at the location with a chosen density level $\rho_c$.

The dense layer advances ($c>0$) at $\mu>0$. This causes the chemical potential to drop at  at locations with lower density ahead of the propagating interface, thereby effectively slowing down the advance of the dense layer. A sharp drop in the dilute layer, leading to a divergent friction factor (\ref{lam1s}), causes substantial deviations from the zero-order density profile, which will be taken into account  in the next section.

\subsection{Evaporation flux  \label{krv}}

We shall consider the case $\rho(\infty) \to 0$, in view of a strong viscous resistance that has lead to the divergence of the effective viscosity in Eq.~(\ref{lam1s}). Any vertical flux causes in this case a substantial change of the chemical potential in the vertical direction, as well as a substantial distortion of the vertical structure of the density field, as will be shown below. 

In a one-dimensional setting, when the flux $j$ is constant, the vertical structure is computed by solving simultaneously the vertical component of the Stokes equation together with Eq.~(\ref{phf0}). We assume $\widehat \eta=$ const (which is justified for the dilute phase, where the friction is most important) and denote $\epsilon =\widehat \eta j$. Using $\rho$ as the independent variable and denoting $\varphi(\rho)=\rho_z^2$, this equation can be rewritten as 
\begin{equation}
\mu(\rho) = g(\rho) - \mbox{$\frac{1}{2}$}\varphi_\rho.
\label{murho}  \end{equation}
This relation can be used in the second Eq.~(\ref{cont1d}), yielding a single equation defining the density profile in the presence of evaporation or condensation. The right-hand side of this equation can be transformed by replacing $v_z = j \,d\rho^{-1} /dz = j \rho^{-2} \sqrt{\varphi}$; an apparent change of the sign of the last term is due to the fact that $\rho_z $ is negative, and has to be defined as $-\sqrt{\varphi}$. The resulting equation can be integrated once, yielding, after some algebra,
\begin{equation}
\rho^2 \frac{d}{d \rho }\left( \frac{\varphi}{ 2\rho} -  f'(\rho) \right)
+ \epsilon \, \frac{\sqrt{\varphi}}{\rho^2} = \alpha.
\label{vrho}  \end{equation}

The integration constant $\alpha$ can be computed by applying this relation deep in the dense layer where $\varphi = \rho_z^2$ vanishes. This gives $\alpha = - \rho_l^2 f'(\rho_l)  \equiv - p(\rho_l)$. Thus, $\alpha$ is identified with the reverse pressure in the bulk of the liquid phase,  but is not defined numerically as yet, since one has still to compute the shift of the liquid density $\rho_l$ from its standard value $\rho_l=1$ due to a shift of the chemical potential $\mu$. Assuming $\mu \ll 1$, one can see that both $\rho_l $ and $\alpha$ are of the same order of magnitude in the bulk of the liquid. One can also observe that, as expected, $\alpha >0$ at $\rho_l <1$ when the dense layer recedes (evaporates). Deep in the vapor phase, one should set $\mu=0$, so that the standard vapor density is not affected. Nevertheless, if the vapor density tends to zero, the boundary condition at the vapor end cannot be applied in a straightforward way, since the term containing $\epsilon$ in Eq.~(\ref{vrho}) is indefinite. The solution strategy can be outlined then as follows. Picking a certain value of $\alpha$, we integrate Eq.~(\ref{vrho}) numerically and compute $\mu(\rho_l)$ using Eq.~(\ref{murho}); then a new value of $\alpha$ is computed with the help of the algebraic equilibrium relations for the liquid phase, and the computation is repeated until it converges to a self-consistent solution.

This procedure can be improved, keeping in mind that both $\mu$ and $\epsilon$ are small, though the relation between them, crucial for our theory, is still unknown. Equation (\ref{vrho}) divided by $\rho^2$ can be formally integrated once more and rewritten in the form
\begin{equation}
\mbox{$\frac{1}{2}$}\varphi -\rho f(\rho)+\rho \int_0^\rho K(\rho')d\rho' = 0
\label{vrho1}  \end{equation}
where
\begin{equation}
K(\rho) =  \epsilon \rho^{-4}\sqrt{\varphi(\rho)} - \alpha  \rho^{-2}.
\label{vrhk}  \end{equation}
Differentiating Eq.~(\ref{vrho1}) yields, in view of Eqs.~ (\ref{gf}) and (\ref{murho}),  
\begin{equation}
\mu (\rho) = \rho K(\rho) + \int_0^\rho K(\rho')d\rho' .
\label{vrho1mu}  \end{equation}
The integral accumulates in the ``boundary layer'' at $\rho \to 0$ where both terms in Eq.~(\ref{vrhk}) diverge; thus, computing $\varphi(\rho)$ in this region is crucial. The scaling in the boundary layer is fixed by requiring all terms in Eq.~(\ref{vrho}) to be of the same order of magnitude. The small parameter $\epsilon$ can be, indeed, eliminated by setting
\begin{equation}
\rho = r \epsilon^{1/3}, \;\; \varphi = \Phi \epsilon^{2/3},\;\;  \alpha=  A \epsilon^{2/3}.
 \label{bscale}  \end{equation}
The rescaled form of Eq.~(\ref{vrho1}) applicable in the boundary layer is, in the leading order,
\begin{equation}
\frac{d}{d r }\left( \frac{\Phi }{r}\right) - 1
+ \frac{2 \sqrt{\Phi}}{r^4} - \frac{2A}{r^2} =0.
\label{vrho0}  \end{equation}

\begin{figure}[tb]
\psfig{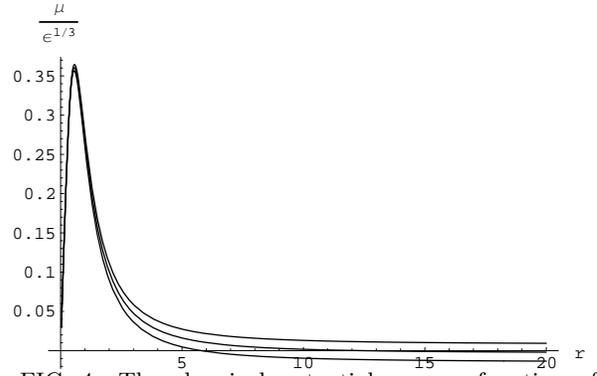}
\caption{The chemical potential $\mu$ as a function of the rescaled density $r$ in the case of evaporation ($\epsilon>0$). The curve corresponding to $A=0.677$ with the asymptotic value of $\mu(r)$ vanishing at $r \to \infty$ is flanked by two curves with positive and negative values of $\mu( \infty)$. \label{fmu}}
\end{figure}

Applying the same scaling to Eq.~(\ref{vrho1mu}), one can see, however, that the generic estimate is $\mu =  O(\epsilon^{1/3})$, which is inconsistent with the equilibrium relationships in the bulk of the liquid. This can be repaired by adjusting $\alpha$ in such a way that the asymptotic value of $\mu$ at $r \to \infty$ vanishes in the leading order. This is, indeed, possible, as proved by integrating Eq.~(\ref{vrho0}) numerically. The integration starts at some $r \ll 0$ using the asymptotic condition $\Phi(r)=A^2 r^4$ at $r \to 0$. A few sample curves $\mu(r)$ at different values of $A$ are drawn in Fig.~\ref{fmu}; all of them approach at $r \to \infty$ at a certain asymptotic value which has to be identified with $\mu(\rho_l)$. The asymptotic value vanishes at $A \approx 0.677$. The residual $O(\epsilon^{2/3})$ value of $\mu(\rho_l)$ satisfying the equilibrium relation $\mu(\rho_l)=\alpha$, can be obtained in the next order by allowing an $O(\epsilon)$ deviation of $\alpha$ from the chosen value $\alpha_c  \approx 0.677\epsilon^{2/3}$. 

Now the solution is completely specified. The chemical potential peaks sharply in the dilute phase (Fig.~\ref{fmu}). This is due to the constraint imposed by a constant flux in one dimension, requiring a large driving force in the transitional layer where a large velocity gradient is necessary to compensate the decreasing density. The asymptotics $d\rho/dz = -\sqrt{\varphi}= - \alpha \rho^2$ corresponds to a rather slow density drop-off, $\rho \propto z^{-1}$ -- a dramatic change, compared to the exponential decay in Eq.~(\ref{phf0}).

The relation between the chemical potential $\mu(\rho_l)$ and flux $j=\widehat \eta^{-1}\epsilon$ obtained above can be reverted, identifying $\mu(\rho_l)=\mu_c$ with the constant chemical potential in the dense layer driving the mass flux $j \approx (0.68 \widehat \eta)^{-1} \mu_c^{3/2}$. In the case of a steadily propagating interface, this can be rewritten as a relation between $\mu_c$ and the propagation velocity $c=-j$; unlike Eq.~(\ref{lam1s}), this relation is now nonlinear:
\begin{equation}
c  = - (0.677 \widehat \eta)^{-1}\mu_c^{3/2} .
\label{lam1sv}  \end{equation}

\subsection{Condensation flux  \label{krv1}}

The above computation is valid only in the case of evaporation ($j>0$ or $c<0$). A constant condensation flux ($j<0$) is, clearly, incompatible with vanishing vapor density. A positive value of $\mu$ is required to enable condensation, i.e.\ advance of the dense phase. This, in turn, implies a finite vapor density, so that $\mu(\rho_v) \approx \rho_v$ at $\rho_v \ll 1$. 

The condensation flux strongly depends on this residual density. Equation (\ref{vrho}) is retained with the sign of $\epsilon$ inverted, but, since the expression multiplying $\epsilon$ is no more indefinite at $z \to \infty$, the integration constant $\alpha$ can be directly related to $\rho_v$:
\begin{equation}
\rho^2 \frac{d}{d \rho }\left( \frac{\varphi}{ 2\rho} -  f'(\rho) \right) +\rho_v^2 f'(\rho_v)
- \epsilon \, \frac{\sqrt{\varphi}}{\rho^2} = 0.
\label{vrhoc}  \end{equation}
The appropriate rescaled variables are again given by Eq.~(\ref{bscale}), and the rescaled  equation replacing Eq.~(\ref{vrho0}) reads, in the leading order
\begin{equation}
\frac{d}{d r }\left( \frac{\Phi }{r}\right) - 1
- \frac{2 \sqrt{\Phi}}{r^4}+ \frac{\beta^2}{r^2} =0,
\label{vert1}  \end{equation}
where $\beta=|\epsilon|^{-1/3}\rho_v$. The asymptotics at $z \to \infty$ or $r \to \beta$ is
 \begin{equation}
\Phi =\kappa^2 (r-\beta)^2,  \;\;\; \kappa=\frac{1}{2\beta^3} \left(1-
{\sqrt{1+4\beta^6}} \right).
\label{vert1a}  \end{equation}
This asymptotics corresponds to an exponential decay of density to its equilibrium value, $r - \beta  \sim e^{\kappa z}$ at $z \to \infty$. 

\begin{figure}[tb]
\psfig{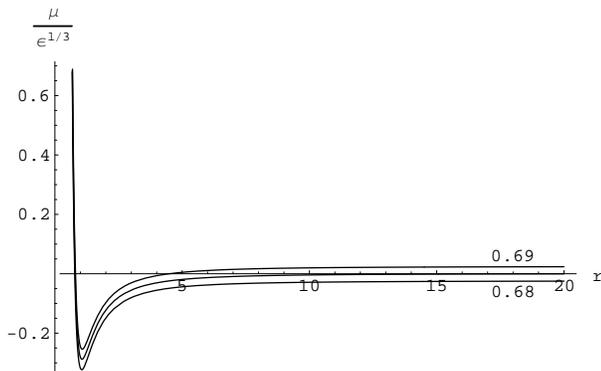}
\caption{The chemical potential $\mu$ as a function of the rescaled density $r$ in the case of condensation ($\epsilon<0$). The curve corresponding to $\beta=0.685$ with the asymptotic value of $\mu(r)$ vanishing at $r \to \infty$ is flanked by two curves with positive and negative values of $\mu( \infty)$. \label{fmu1}}
\end{figure}

Integrating Eq.~(\ref{vert1}) with the asymptotic condition (\ref{vert1a}), one can see that also in this case the chemical potential defined by Eq.~(\ref{murho}) reaches a constant asymptotic value at $r \to \infty$. Checking the asymptotics of Eq.~(\ref{vrhoc}) at $\rho \to \rho_l$, one can see that the thermodynamic pressure on the liquid side, $\rho_l^2 f'(\rho_l)$ should be equal that on the vapor side, $\rho_v^2 f'(\rho_v)$, and, hence, be of $O(\epsilon^{2/3})$. This is again inconsistent with the generic estimate $\mu =  O(\epsilon^{1/3})$, implying, through the equilibrium relationships, the same order of magnitude of $f'(\rho_l)$. Hence, as in the preceding subsection, the value of $\beta$ has to be adjusted in such a way that the asymptotic value of $\mu$ at $r \to \infty$ vanishes in the leading order. The value found by shooting is $\beta \approx 0.685$ (see  Fig.~\ref{fmu1}).

\subsection{Kinetically retarded motion  \label{kret}}

Taking into account ``normal'' viscous retardation only (with $\widehat \eta \approx \eta$) may exaggerate the actual phase transition rate, since transport through a sharp density gradient is in fact an activated process, except, perhaps, in an immediate vicinity of a critical point. When the interface is treated as a sharp discontinuity, this may be accounted for by introducing a finite evaporation rate (involving an appropriate activation energy) and a condensation ``sticking coefficient''. Both quantities are difficult to estimate quantitatively but, in principle, they insure a finite evaporation or condensation rate even under conditions when viscous retardation is absent.

In the framework of the phase field theory, kinetic retardation can be accounted for by replacing the stationary equation (\ref{phf}) or (\ref{phf0}) by the respective gradient flow equation containing a large relaxation time $\tau$. In one dimension, we have
\begin{equation}
\tau \rho_t = \rho_{zz}- g(\rho) + \mu.
\label{phft}  \end{equation}

On the infinite axis, this equation (with $\mu =$ const) has a solution steadily propagating with a speed dependent on $\mu$, and satisfying the stationary equation in the comoving frame
\begin{equation}
- \tau c \rho_z+ \rho_{zz}- g(\rho) + \mu=0.
\label{phfc}  \end{equation}

In the case of weak disequilibrium, $\mu=O(\delta) \ll 1$, the propagation speed $c =O(\delta)$ is easily computed, as in the preceding subsection, using the solvability condition of Eq.~(\ref{phfc}) expanded in $\delta$:
\begin{equation}
c = \frac{ \mu (\rho_l - \rho_v)}{\tau \sigma}
 =\frac{6 \mu}{\tau },
\label{csolv}  \end{equation}
where $\sigma$ is defined by Eq.~(\ref{sig}) and the numerical value is given for the cubic $g(\rho)$.

Equations (\ref{lam1s}) and (\ref{csolv}) represent two opposite limits when, respectively, either viscous or kinetic retardation is prevalent. A rough estimate for the lower bound of the relaxation time is $\tau \propto l^2/D$, where $l$ is the thickness of the diffuse interface and $D$ is the diffusivity. The characteristic time of viscous retardation on the same length scale is $\tau_v \propto l^2/\nu$, where $\nu=\eta/\rho$. For common liquids, the Prandtl number Pr $=\nu/D$ is large, and $\tau_v/\tau \propto D/\nu \ll 1$. Viscous retardation may be still felt at larger scales, complementing the kinetic retardation near the diffuse boundary. At Pr $\gg 1$, the flow velocity is nearly constant throughout the transitional boundary region, and the propagation velocity defined by Eq.~(\ref{csolv}) can be viewed as the velocity of the slow drift of the interphase boundary due to the evaporation or condensation in the frame moving with the local velocity of the ambient fluid. At fixed propagation velocity, the increments due to the viscous and kinetic retardation are additive. In the dense layer, the former is negligible at Pr $\gg 1$, although it becomes important in the dilute phase, as we have seen in the preceding subsections. 

The scaling of the lubrication approximation (Section \ref{curved}) remains consistent only if the relaxation time $\tau$ in Eq.~(\ref{phft}) is of $O(\delta)$. With this scaling, the speed of the vapor-liquid interface displacement is of $O(\delta^2)$, i.e. of the same order of magnitude as the vertical velocity.

\subsection{Spreading assisted by interphase transport  \label{krh}}

The results of the computations that have been carried out so far in this section for an infinite fluid layer separated by a diffuse vapor-liquid interface can be applied to the spreading problem after minimal modification. In a bounded layer, the chemical potential in the liquid phase $\mu_c$ driving the evaporation or condensation flux is determined by the combined action of surface tension and disjoining pressure. The disjoining potential can be computed with the help of the solvability condition, as in Section \ref{S1ds}; the respective formulae remain in force, since the flux-related drop of the chemical potential occurs in the dilute phase only, and is negligible in the diffuse interface region, where the translational eigenfunction is localized. 

The basic equation of the lubrication approximation, Eq.~(\ref{hs}), is modified in the case of non-equilibrium spreading by an added evaporation or condensation term: 
\begin{equation}
\frac{\partial h}{\partial t} = j(P) + \eta^{-1} \partial_x \left [ Q(h) P_x \right] .
\label{phft1}
\end{equation}
The expressions for the evaporation or condensation flux $j(P)$ and even the orders of magnitude vary depending on the physical situation under consideration, according to the calculations presented in the three preceding subsections, with the effective pressure defined by Eq.~(\ref{stoxp}) replacing $\mu_c$. This determines, in turn, the relative importance of the two terms on the right-hand side of Eq.~(\ref{phft1}). In the case of viscously retarded motion with finite $\rho_v$, $j$ happens to be proportional to $ \widehat \eta P$, although the term representing the horizontal transport through the liquid phase is of order $ \widehat \eta\ \partial^2P/\partial x ^2$, and is negligible compared to the evaporation or condensation term in the lubrication limit, when the horizontal derivatives are small. In this case, flow across the isodensity levels associated with evaporation or condensation, driven by the deviation of the chemical potential from equilibrium, would be larger by $O(\delta^{-1})$ than hydrodynamic ``horizontal'' motion. Therefore, it is likely that this does not represent the most usual situations, where evaporation is hindered by large activation energies. In the present model a way to enter consistently this activation effect is to make the evaporation flux and the horizontal transport of the same of magnitude. This is done by imposing an $O(\delta^{-1})$ relaxation time $\tau$. Although this connection between a molecular quantity $\tau$ and a macroscopic length scale $\delta$ may look a bit artificial, this represents a distinguished limit where a balance between factors of different physical origin is attained. 

The last situation that we have to consider is the one of a vapor phase of vanishingly small density when the evaporation flux is related to the jump of chemical potential as $j \propto \mu_c ^{3/2}$. In this case, it is possible to have the evaporation flux and the horizontal transport of the same order of magnitude with the choice of scaling $\tau =O(\delta^{-1/2 })$. A slow density decay caused by evaporation, which might lead to a weakly (logarithmically) divergent horizontal flux, may be a disturbing factor, but this is certainly an artifact caused by a constant flux in a one-dimensional setting and not transferable to two-dimensional spreading.  It is of interest to notice at this point that, when $j$ is dominant, and in the absence of horizontal flux (which would happen far from a solid boundary), one recovers the classical Thomson expression for the evaporation driven by the curvature of the liquid-vapor interface.

Mass transport across isodensity lines should become particularly important when the lubrication approximation breaks down. This should happen near the ``contact line'' in the case when two alternative fluid densities near the solid wall are possible (see Section \ref{S1bc}). If, say, the boundary densities are $\rho_{sv} \ll 1$ and $\rho_{sl}=1-a, \; a \ll 1$, the three-phase ``contact line'' can be viewed as a sharp transition between $O(1)$ positive and negative values of the nominal thickness $h$, such that $e^{-|h|} \ll 1$ on either side. This can be treated as a shock of Eq.~(\ref{hs}) or (\ref{hs1}). The Hugoniot condition which should ensure zero net flux through the shock is the equality of chemical potentials on both sides. Unfortunately, this condition cannot be formulated precisely, since the sharp-interface limit of the surface tension term is inapplicable in the shock region. Moreover, our test computations of the profile of the dense layer using Eq.~(\ref{hsy}) with different boundary conditions imposed on the ``shock'' at $h=h_0$ showed that the spreading velocity is very sensitive to the conditions on the shock. 

It remains therefore essential to solve the full system of density field and hydrodynamic equation in the shock region whenever a sharp transition between alternative surface densities is possible. The outer limit of the resolved shock structure should be matched with the lubrication equations (\ref{hs}), (\ref{hs1}), or (\ref{phft1}). The transport across isodensity lines in the shock region alleviates the viscous stress singularity remaining in the lubrication model. In its turn, the latter provides a gradual transition to the  sharp-interface limit at large distances.  

\section{Summary and perspectives}

As well known, the phase field model provides a sound theoretical basis for studying equilibrium capillary phenomena in fluids.  It allows to derive in a straightforward manner the classical formulae for the capillary pressure and for the equilibrium contact angle, contrary to formulations based upon the introduction of van der Waals forces diverging at short distances.  We have shown that this model can be extended in a natural way to study a thoroughly dynamical spreading process.  The lubrication limit, where the contact angle is small, allows to derive consistently an equation of motion for the liquid-vapor interface interacting with the solid surface.  In the static limit, this equation yields back the equilibrium Young-Laplace theory.  

Evaporation or condensation are processes that are included in this model.  The driving force  for the evaporation or condensation is the imbalance between the pressure drop across the interface and its equilibrium value.  Similarly, and consistently with Seppecher's \cite{sepp} results, an advancing or receding contact angle differing from its equilibrium value makes the contact line a source or sink for evaporation or condensation. We suggest to check experimentally this interesting phenomenon by observing the accumulation of a non-volatile tracer diluted in the liquid phase that would be left by evaporation near a moving contact line. 

Our analysis indicates that kinetic retardation of the interphase transport is essential for a well balanced theory away from the critical point. The available simulations of the motion of a diffuse interface near a three-phase contact line \cite{sepp,jacq}, taking into account viscous retardation only and, in effect, assuming evaporation or condensation to be as easy as plain advection, may grossly overestimate the rate of interphase transport, but the latter remains essential even when its order of magnitude is reduced due to kinetic retardation.

The present theory extends itself in a very natural way to problems like film breaking. The latter situation is interesting also because it should allow to approach experimentally thermodynamical critical points, where phase field models certainly apply, although things should become complicated if the solid-fluid interaction is added to the critical phenomena near a moving contact line. 

\acknowledgments
This work has been supported by the Israel Science Foundation and the EU TMR network ``Nonlinear dynamics and statistical physics of spatially extended systems''. A part of this work was carried out during the stay of LMP at the Laboratoire de Physique Statistique de l'Ecole Normale Sup\'erieure supported by CNRS, the visit of both authors to the Instituto Pluridisciplinar, Universidad Complutense, Madrid, and the visit of YP to the Technion; hospitality of all these institutions is acknowledged.

\end{document}